\documentclass[aps,prl,floats,floatfix,twocolumn]{revtex4}

\usepackage{ifthen}
\usepackage{ifpdf}
\usepackage{color}

\ifpdf
\usepackage{graphicx}
\usepackage{epstopdf}
\else
\usepackage{graphicx}
\usepackage{epsfig}
\fi
\graphicspath{{./Figs/}{./}}

\usepackage{latexsym}
\usepackage{amsmath}
\usepackage{amssymb}
\usepackage{bm}
\usepackage{wasysym}

\newcommand{\amatrix}[1]{\begin{matrix} #1 \end{matrix}}

\newcommand{\braket}[1]{ \left\langle #1 \right\rangle}

\newcommand{\be}[1]{\begin{eqnarray}\ifthenelse{#1=-1}{\nonumber}{\ifthenelse{#1=0}{}{\label{e#1}}}}
\newcommand{\beq}{\begin{eqnarray}}
\newcommand{\eeq}{\end{eqnarray}} 

\newcommand{\hide}[1]{}
\newcommand{\Eq}[1]{\textcolor{blue}{{Eq.}\!\!~(\ref{#1})}}

\newcommand{\Fig}[1]{\textcolor{blue}{Fig.}\!~\ref{#1}}
\newcommand{\sect}[1]{{\bf #1.-- }}

\renewcommand{\cite}[1]{\textcolor{blue}{[\onlinecite{#1}]}} 

\begin{document}

\title{Hidden timescale in the response of harmonically driven chaotic systems}

\author{Christine Khripkov$^1$, Amichay Vardi$^1$, and Doron Cohen$^2$}

\affiliation{
\mbox{$^1$Department of Chemistry, Ben-Gurion University of the Negev, Beer Sheva 84105, Israel}
\mbox{$^2$Department of Physics, Ben Gurion University of the Negev, Beer Sheva 84105, Israel}
} 

\begin{abstract}
Linear response theory relates the response of a system to the power-spectrum of its fluctuations. However, the response to external driving in realistic models exhibits a pronounced non-linear blurring of the spectral line-shape. Considering a driven Bose-Hubbard trimer model we figure out what is the hidden time scale that controls  this smearing effect. Contrary to conventional wisdom, the Fermi-golden-rule picture fails miserably in predicting the non-linear width of the transitions. Instead, if the system has a classical limit, the determination of the hidden time scale requires taking into account the underlying classical phase-space dynamics.   
\end{abstract}
\maketitle


The response 
of a system to a driving source~$f(t)$ is a recurring major theme in mechanics. 
The resulting diffusive spreading of the system's energy distribution is often treated within the well known framework 
of linear response theory (LRT) with  
its celebrated fluctuation-dissipation relation.   
From a mesoscopic perspective, the derivation 
of the Kubo formula for the diffusion in energy 
requires the assumption of chaotic ergodicity \cite{ott1,ott2}.
In a quantum context, an attempt has been made
to extract LRT from the Fermi-golden-rule (FGR) picture 
of transitions between levels \cite{wilk0,Wilk}.
However, this derivation turned out to be non-trivial \cite{WilkA},  
and has motivated more elaborate studies \cite{crs,rsp,frc}.
Considering a driven chaotic system,
\be{1}
{\cal H} \ \ = \ \ {\cal H}_0 +f(t)W
\eeq
with a weak noisy perturbation $f(t)$, it is claimed 
that the transition rate from level~$m$ to level~$n$ is  
\be{2}
w_{nm} \ \ = \ \ |W_{n,m}|^2 \ \mathcal{A}^2 \ 2\pi \tilde{F}(E_n{-}E_m) 
\eeq
where $\mathcal{A}^2\tilde{F}(\omega)$ is the power spectrum
of $f(t)$, and $\mathcal{A}$ is its RMS value. If the correlation time of the noisy perturbation 
is~$\tau_{\text{noise}}$,  then $\tilde{F}(\omega)$ has $1/\tau_{\text{noise}}$ width.
However, for the very common case of harmonic driving ${f(t)=A\sin(\Omega t)}$ 
(such that ${\mathcal{A}=A/\sqrt{2}}$), ${\tilde{F}(\omega)}$ is a sum of delta functions, 
namely ${(1/2)\sum\delta(\omega{\pm}\Omega)}$. 
The standard phenomenology is to assume 
that any finite amplitude~$\mathcal{A}$ implies
an intrinsic width~$1/\tau$ that depends on~$\mathcal{A}$.
Thus the delta functions are {\em broadened}, 
\beq
\delta(\omega-\Omega) \ \ \mapsto \ \ \tau \,  G\left(\frac{\omega-\Omega}{1/\tau}\right) 
\label{bdf}
\eeq  
where $G(x)$ is a normalized Gaussian-like function 
(the exact form of $G(x)$ is practically insignificant).
We argue below that an appropriate prescription is in fact
\be{3}
\tilde{F}(\omega) \ = \
\left[\frac{1}{1+(\Omega\tau)^{-2}}\right] \frac{\tau}{2} \sum_{\pm}  G\left(\frac{\omega\pm\Omega}{1/\tau}\right) 
\eeq
Without the expression in the square brackets, 
or for $\Omega\tau\gg 1$,  
it coincides with the naive procedure
of \Eq{bdf}.

The intrinsic time scale~$\tau$ does not manifest itself in the traditional LRT analysis:
it cancels out in any formal calculation of energy spreading under the standard assumption that  the width~$1/\tau$ 
is much larger than the level spacing, but much narrower than any other spectral feature. 
In idealized circumstances one assumes {\bf (a)}~an homogeneous density of levels~$g(E)$, 
and {\bf (b)}~a flat band. The latter requirement means that the squared matrix elements $|W_{nm}|^2$ 
are statistically independent of~$E_n$ and~$E_m$ in the energy range of interest. 
Such idealization underlies the Random Matrix Theory (RMT) approach of Wigner and followers. 
The outcomes of LRT is the Kubo expression for the diffusion coefficient, 
which can be written schematically as ${D_E=c(\Omega)\mathcal{A}^2}$.  
Within this framework, the FGR-based {\em quantum version} of the Kubo formula gives  
the same result as the {\em classical version}, up to ``weak" corrections \cite{lrt}. 
This is somewhat analogous to the Thomas-Reiche-Kuhn \mbox{$f$-sum-rule}, 
and has been termed {\em restricted} quantum-classical correspondence (restricted QCC).

The flat-band assumption is generally not satisfied in mesoscopic systems of physical interest.
Thus, in general we expect the hidden time scale~$\tau(\mathcal{A})$ 
to manifest itself in any realistic energy spreading process. It is therefore implied that 
the Kubo expression for the diffusion in energy acquires a non-linear 
dependence, namely, 
\be{4}
D_E \ \ = \ \ c(\Omega;\tau(\mathcal{A})) \ \mathcal{A}^2
\eeq
In the present work we offer predictions for the functions~$c(\Omega,\tau)$ and~$\tau(\mathcal{A})$, 
and test them on a concrete and experimentally-realizable model system.  Furthermore, we would like to determine whether the evaluation of~$\tau(\mathcal{A})$ requires quantum mechanics, 
or maybe QCC holds with regard to this hidden time scale too.

\sect{Model system}
We consider a Bose-Hubbard trimer \cite{trimer1,trimer2,trimer3,trimer4,trimer5,trimerD1,trimerD2,trimerE} 
with $N$ particles, 
\be{55}
\mathcal{H} \ \ = \ \
\frac{U}{2}\sum_{j=0}^2 n_j^2 
\ + \ \frac{K}{2} \sum_{j=1}^{2} \left(a_j^{\dag} a_0 + a_0^{\dag}a_j\right) 
\label{BHT}
\eeq
Here ${j=0,1,2}$ labels the three modes, $a_{j}$ and $a_{j}^{\dagger}$ are canonical destruction and creation operators in second quantization, $K$ is the hopping frequency, and $U$ is the on-site interaction. This system is the minimal Bose-Hubbard model admitting chaos. The dynamics of the undriven system at any given energy $E$ is determined by a single dimensionless parameter $u=UN/K$. The classical limit is attained as $N\rightarrow\infty$ while $u$ is kept constant. At this limit quantum fluctuations diminish and the field operators $a_i$ may be replaced by $c$-numbers. We identify the chaotic regions  in the $\{E,u\}$ parameter space from the quantum level spacing statistics, using either the Brody-parameter map \cite{Brody81} or the adjacent level spacing correlation function \cite{Oganasyen07}, as detailed in references \cite{Tikhonenkov13,Khripkov15b}. Further verification of chaoticity is obtained from the classical Poincare sections at the same parameter values. Below we take $u=3$ and initiate the system in the middle of the energy spectrum, i.e. $\varepsilon(t=0)=1/2$ where $\varepsilon\equiv E/(E_{\rm max}-E_{\rm min})$ and $E_{\rm min,max}$ denote the extremal energies of the spectrum at the pertinent value of $u$. This choice ensures the chaotic dynamics of the undriven trimer system.

The trimer of \Eq{e55} is subjected to weak harmonic driving via its hopping term, as in \cite{Tikhonenkov13}, 
namely, ${(K/2) \mapsto (K/2) + A\sin{\Omega t}}$, and
hence the perturbation term in \Eq{e1} is $W=\sum_{j=1}^{2} \left(a_j^{\dag} a_0 + a_0^{\dag}a_j\right) $.

\sect{Band profile}
The classical power spectrum of the perturbation~$W$ in the absence of driving 
can be obtained by calculating the Fourier transform of $\braket{W(t)W(0)}$ 
for a long ergodic trajectory. It corresponds to the quantum spectral function
\beq
\tilde{C}(\omega) \ = \ \sum_{n(\neq m)} \left | W_{nm}\right|^2 2\pi\delta\left(\omega-\left( E_n-E_m\right)\right)
\label{cqm}
\eeq
with an implicit averaging over the reference state~$m$ 
within the energy window of interest.
A rough but pedagogical way to write this formula 
is ${\tilde{C}(\omega)=2\pi g(E) |W|^2}$
where $g(E)$ is the density of states.  
It shows that the power-spectrum reflects 
the band-profile of the perturbation matrix. 

\begin{figure}

\includegraphics[width=\hsize]{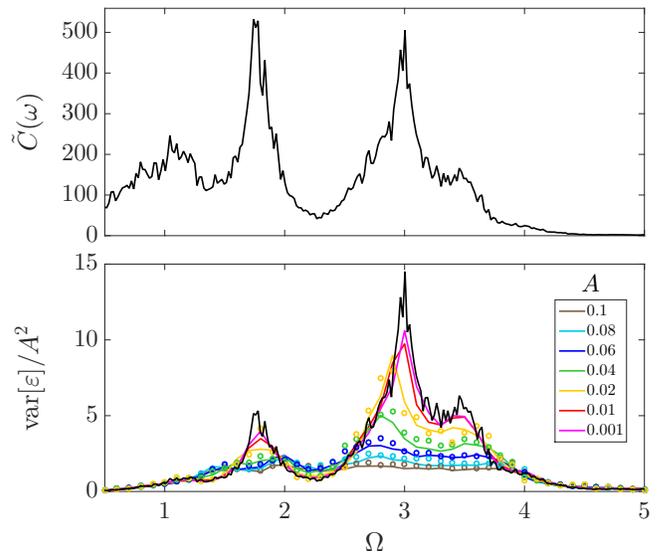}

\caption{(color online) 
The classical power spectrum of the fluctuating driving force (upper panel), 
and the frequency dependence of the instantaneous energy variance at ${Kt=50}$ 
for various drive intensities (lower panel).
Units of time are chosen such that ${K=1}$. 
Results for both classical (solid lines) and quantum (markers) simulations are presented. 
The black solid line in the lower panel corresponds to 
the standard LRT prediction $D_E \approx \Omega^2 {\tilde C}(\omega=\Omega)$.
}

\label{f1}
\end{figure}

Common random matrix models assume that the band profile is flat, in the sense
that a band ${|\omega|<\Delta_b}$ can be defined, within which all 
matrix elements are comparable in size. The bandwidth for strongly chaotic systems 
is related to the classical correlation time, 
namely ${\Delta_b=2\pi/\tau_{cl}}$. However, the classical power spectrum  of the Bose-Hubbard trimer (top panel of  \Fig{f1}) is by no means flat, because the model system 
is not strongly chaotic. Thus, the correlation time and the bandwidth are quantitatively ill-defined.   
This is in fact the typical situation for any realistic non-artificial model of physical interest.

\sect{Transition rates for low drive frequencies}
The FGR 
implies that the transition rate from an initial level~$E_m$ 
to another level $E_n$ is ${w = 2\pi |W|^2 \mathcal{A}^2 \tau}$,
where $\mathcal{A}^2 \tau$ and $|W|^2$ are the intensity 
of the driving source and the associated coupling strength  
for the pertinent transition frequency.  While this estimate is valid in the high frequency range
 $\Omega\gg 1/\tau$,  we argue below that for the purpose of energy-spreading analysis 
this rule should be extended as:
\be{5}
w \ \ = \ \ 2\pi |W|^2 \times \left\{ \amatrix{
(\mathcal{A}\Omega)^2 \tau^3 & \text{for} \ \Omega \tau \ll 1 \cr
\mathcal{A}^2\tau & \text{for} \ \Omega \tau \gg 1
}
\right.
\eeq  
Interpolation between the low-frequency
("DC") and high-frequency ("AC") regimes,  then
leads to the expression in the square brackets of \Eq{e3}.
Note that the normalization 
of ${\mathcal{A}^2\tilde{F}(\omega)}$ in the AC regime 
reflects the variance of the amplitude ${\text{Var}[f]=\mathcal{A}^2}$, 
while in the DC regime it reflects 
the variance of the sweep rate ${\text{Var}[\dot{f}]=(\mathcal{A}\Omega)^2}$.  

In order to understand the DC extension of \Eq{e5}, consider
first a flat-band within the energy range ${|E_n-E_m-\Omega|\lesssim (1/\tau)}$. 
This band contains ${\mathcal{N} \approx (1/\tau)g(E)}$ levels.
The standard AC expression implies that the 
total FGR rate 
\beq
\Gamma \ \ = \ \ \mathcal{N}w \ \ = \ \ 2\pi \mathcal{A}^2 |W|^2 g(E)
\eeq
is independent of $\Omega$. This is clearly {\em false} in the DC regime,  
because the rate of transitions should vanish in the adiabatic limit. 
The proper procedure in the DC regime, 
is to switch from the fixed-basis to the adiabatic-basis 
representation (for details see \cite{crs,rsp,frc}).
Consequently  $\mathcal{A}$ and $W$ are transformed as ${\mathcal{A} \mapsto (\mathcal{A}\Omega)}$ and ${W \mapsto (W/\omega)}$. Thus, with ${\omega \sim (1/\tau)}$ the AC formula 
is replaced by the DC version of \Eq{e5}.  
It is implied that ${\tilde{\Gamma} = (\Omega \tau)^2 \Gamma \propto (A\Omega)^2 }$
is the effective ``level broadening". This DC broadening depends 
on the sweep rate ${|\dot{f}|\sim (\mathcal{A}\Omega)}$, 
in contrast to the AC broadening that depends predominantly on~$\mathcal{A}$.

\sect{The diffusion coefficient}
The diffusion coefficient characterizes the second-moment 
of the spreading process, namely, $(\delta E)^2=2D_E t$. 
In the AC regime it is estimated as ${D_E \sim \Gamma \Omega^2}$, 
while in the DC regime it is estimated as ${D_E \sim \tilde{\Gamma} (1/\tau)^2}$.
Therefore, under a flat-band assumption we get formally the {\em same} expression in 
both regimes, and the $\tau$ dependence cancels out.
 
When $\tilde{C}(\omega)$ is not flat, a more careful calculation is required, accounting for 
the $\omega$~dependence of the rates~$w_{nm}$ in \Eq{e2}. Summing over all the possible 
transitions we obtain \Eq{e4} with
\be{9}
c(\Omega,\tau) \ \ = \ \ \int_0^\infty \omega^2 {\tilde C}(\omega) {\tilde F}(\omega) d\omega
\eeq 
This formula is often used to evaluate the decay rates of systems subjected to noisy perturbations, as in the Zeno and anti-Zeno effects \cite{Kofman00}. It implies $\tau$-sensitivity in realistic models
where the flat-band assumption is inapplicable. 
Hence we anticipate the existence 
of the hidden time scale $\tau$ to be exposed.

\sect{Long-time energy spreading}
Given the diffusion coefficient ${D_E}$ the long time evolution of the system's energy
distribution can be deduced. 
LRT is based on the observation that 
upon coarse-graining the coherent transitions 
become stochastic-like \cite{Wilk,WilkA,crs,rsp,frc,lrt}, 
hence the time evolution of the probabilities ${p_n(t)=|\langle n | \psi(t) \rangle |^2}$ 
obeys a master equation:
\beq
\label{FGR}
\frac{d}{dt} p_n \ = \ \sum_m w_{nm} \left( p_m - p_n\right)~,
\eeq
where the transition rates are given by \Eq{e2}.
From here one deduces the Fokker-Planck equation (FPE) 
for the diffusion in energy space \cite{Jar,Jar2,kafriN,Ates12,Tikhonenkov13,Niemayer13,Niemayer14,Khripkov14a}:
\beq
\frac{d}{dt} \rho(E) \ = \ \frac{d}{dE}\left[ g(E) D_E\frac{d}{dE} \left(\frac{\rho(E)}{g(E)} \right) \right]~,
\label{FPE}
\eeq
where $\rho(E)$ is the coarse-grained density that is associated with~$p_n$, 
and $g(E)$ is the density of states.

\begin{figure}

\includegraphics[width=\hsize]{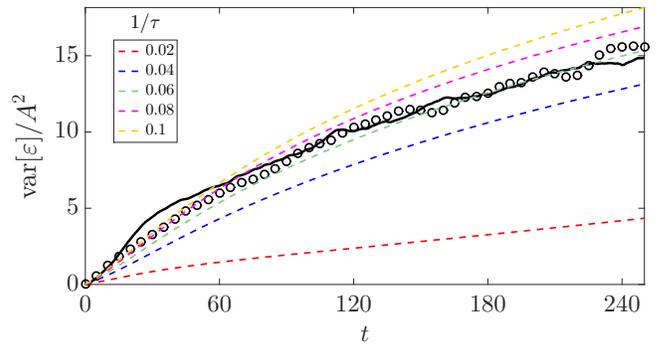} \\

\caption{(color online) 
The time evolution of the classical (solid line) 
and quantum (circles) energy variance is compared with the evolution 
that is generated by the master equation \Eq{FGR} for several values of $1/\tau$ (dashed lines). 
Such plots are used to determine the best-fit $\tau$ via a least-mean-square procedure.
}

\label{f2}
\end{figure}

\begin{figure}

\includegraphics[width=\hsize]{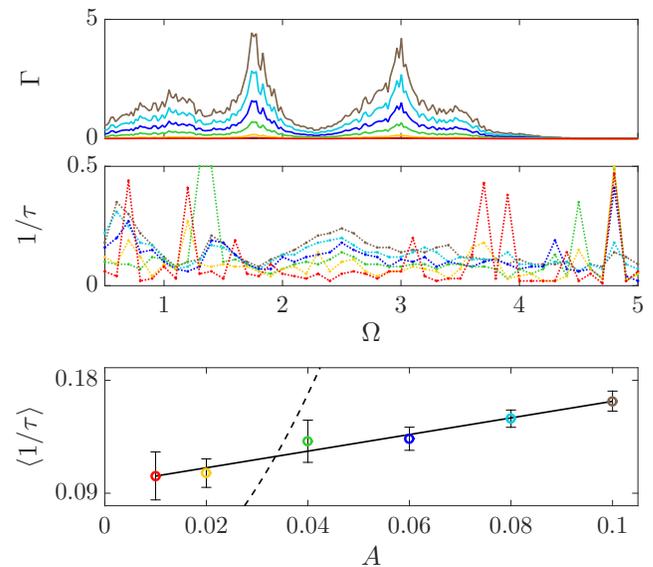} \\

\caption{(color online) 
(a) The FGR width $\Gamma$ is calculated as a function of $\Omega$ 
and displayed for various values of the drive intensity~$\mathcal{A}$.  
(b)~The best fit for the width $1/\tau$, as obtained from plots 
such as \Fig{f2}, is displayed for the same frequencies and amplitudes.
Note that the dependence on $\Omega$ is anticorrelated with respect 
to the FGR width.
(c)~Averaging over $\Omega$ reveals linear dependence 
of $1/\tau$ on $\mathcal{A}$. The dashed line is 
the FGR prediction $1/\tau=\Gamma$ based on the first panel. 
}

\label{f3}
\end{figure}

\sect{Prediction for the hidden time scale}
In order to calculate $D_E$ using \Eq{e9} with \Eq{e3} 
we have to know how $\tau$ depends on~$\mathcal{A}$ and~$\Omega$.
We first attempt to propose a self-consistent prediction:
\be{8}
\frac{1}{\tau} \ \ = \ \ \min\left\{ \ \Gamma, \ \ \left(\Omega^2\Gamma\right)^{1/3}, \ \ cA \ \right\}
\eeq
where $c$~is a constant that is determined by the 
classical dynamics in phase-space. 
The first entry in \Eq{e8} is the naive guess ${(1/\tau)=\Gamma \propto A^2}$. 
However in view of \Eq{e3} the self-consistent equation takes the form 
\beq
\frac{1}{\tau} \ \ = \ \  \left[\frac{1}{1+(\Omega\tau)^{-2}}\right] \Gamma 
\eeq
The solution of this equation in the adiabatic regime  
leads to the second entry in \Eq{e8}, 
that features a slower dependence on~$\mathcal{A}$, 
namely, ${(1/\tau) \propto \mathcal{A}^{2/3}}$. 
 
At this stage one wonders what is the condition for the validity 
of the self-consistent FGR approach. This has been discussed in \cite{crs,rsp,frc}.
The key observation is that the small parameter of the theory
is $(1/\tau)/\Delta_b$ \cite{Ra}.  
However, this condition is very difficult to satisfy for structured $\tilde{C}(\omega)$ 
band-profiles (as in \Fig{f1}) because stretches of flat-band are 
very small, especially in the vicinity of sharp peaks. The FGR picture should therefore 
be supplemented with an additional parameter $\tau$ that cannot be determined 
self-consistently from FGR considerations alone.
  
If the FGR validity condition ${(1/\tau) \ll \Delta_b}$ is violated, 
then ${\tau \ll \tau_{cl}}$. This requirement of 
having non-perturbative mixing of levels prior to $\tau_{cl}$ 
is in-fact a necessary condition for detailed semi-classical correspondence.
If detailed QCC holds, one should expect a leading linear 
dependence ${(1/\tau) \propto \mathcal{A}}$, which is the third entry in \Eq{e8}. 
The reasoning is as follows: for sake of argumentation assume that  $\tilde{C}(\omega)$  
possesses a peak at some frequency~$\omega_r$ that corresponds    
to some classical resonance. If the driving amplitude is~$\mathcal{A}$, 
then the adiabatic energy surface ${\mathcal{H}=E}$ will have 
a variation ${\propto \mathcal{A}}$ in phase space, leading to a smearing 
of the resonance position over an associated frequency scale~${\propto \mathcal{A}}$.
Similar effect can be caused by blurring of regions in mixed phase-space.

\begin{figure}
\includegraphics[width=\hsize]{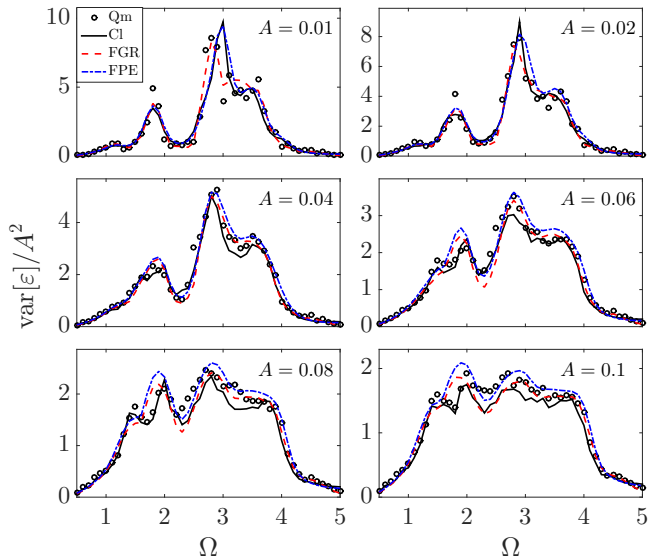}

\caption{(color online) Comparison of the instantaneous variance at ${Kt=50}$ for quantum (circles) and classical (solid line) simulations with the same quantity obtained from the FGR rate equation \Eq{FGR} (dashed line) and the FPE equation \Eq{FPE} (dash-dotted line) using the $\tau(A)$ of \Fig{f2}c.}

\label{f4}
\end{figure}

\sect{Manifestation of the hidden time scale}
We turn to present a numerical procedure for detecting the hidden time scale. 
Our testing ground is the Bose-Hubbard model of \Eq{e55}.
Subjecting the trimer to driving as described above, 
we carry out both classical and quantum propagation 
and follow the time evolution of the expanding energy distribution $p_n(t)$, 
calculating its variance $(\delta E)^2$ at each instant of time.  
In \Fig{f1}b we plot the instantaneous variance as a function of the drive frequency $\Omega$ after a predetermined evolution time.
LRT predicts that the same dependence of $\delta E^2/\mathcal{A}^2$ on the drive frequency $\Omega$ would be obtained for all drive intensities, i.e. $\delta E^2/\mathcal{A}^2\sim\Omega^2{\tilde C}(\omega=\Omega)$ (solid black line). While at low drive intensities this is indeed roughly the case, higher intensities result in broadening of the response profile. We benchmark the theoretical predictions of the previous sections by assessing  whether this non-linear response effect can be reproduced by introducing the hidden time scale $\tau$ into the analysis. 

As demonstrated in \Fig{f2}, the time dependence of $(\delta E)^2 /\mathcal{A}^2$ on $\mathcal{A}$ can be indeed reproduced by fitting the {\em single} parameter $\tau$ in the FGR simulation of \Eq{FGR}. The best-fit  $1/\tau$ values for various values of $\Omega$ (\Fig{f3}b) are much smaller than the naive FGR-based expectation ${(1/\tau) \approx \Gamma \propto \mathcal{A}^2}$ (\Fig{f3}a), indicating the complete  failure of the standard LRT. This is a-priori expected from the former discussion of \Eq{e8}, because the condition ${\Gamma<\Delta_b}$ is largely violated.

The~$\mathcal{A}$ dependence of the frequency averaged~$1/\tau$ is presented in  \Fig{f3}c. We observe that this effective width depends linearly on~$\mathcal{A}$, indicating that the semi-classical perspective (third entry in the r.h.s. of \Eq{e8}) is most appropriate for its determination. For comparison, the naive quadratic LRT prediction (dashed line) is nowhere near the numerical results.

Having obtained in \Fig{f3}c the $\mathcal{A}$ dependence 
of the  single parameter $\tau(\mathcal{A})$, 
we check whether it can be used in order to reproduce
the full $\Omega$ dependence of the spreading variance. 
Using either the FGR rate equation, \Eq{FGR},   
or its coarse-grained FPE version \Eq{FPE}, we show 
in \Fig{f4} that the details of the line broadening 
are accurately reproduced.  The dependence of~$1/\tau$ on $\Omega$ 
can therefore be neglected in practice. 
Still, there is a weak modulation in the middle panel of \Fig{f3} that exhibits anti-correlation 
with the $\Gamma$ of the top panel. We attribute this modulation to a residual FGR 
effect implied by the discussion of \Eq{e8}: wherever the FGR condition is not violated, 
it predicts a larger ${1/\tau}$ compared with the linear semi-classical estimate.

\sect{Conclusions}
The analysis and numerical results of this work demonstrate that 
the nonlinear smearing 
of the LRT response line-shape can be explained 
by introducing an effective $\mathcal{A}$-dependent width ${1/\tau}$
in the expression for the FGR transition rates. 
The determination of the hidden time scale $\tau$ 
requires to go beyond the conventional FGR picture.

{\bf Acknowledgments.-- }
This research was supported by the Israel Science Foundation (grant Nos. 346/11 and 29/11).


\clearpage

\begin{thebibliography}{99}
\bibitem[a]{Ra}
In the adiabatic regime ${[(1/\tau)/\Delta_b]^{3/2} \ll 1}$ 
provides the slowness condition on the sweep rate~${|\dot{f}| \sim  \mathcal{A}\Omega}$, 
while in the ``AC" regime the sensitivity is mainly to~$\mathcal{A}$.
\end{thebibliography}
\end{document}